\documentstyle[11pt]{article}

\input epsfig.sty

\renewcommand{\thefootnote}{\fnsymbol{footnote}}

\begin{document} \begin{titlepage} 
\rightline{\vbox{\halign{&#\hfil\cr
&SLAC-PUB-7373\cr
&UTEXAS-HEP-96-21\cr
&DOE-ER-40757090\cr
&December 1996\cr}}}
\vspace{0.35in} 
\begin{center}

{\Large\bf
Probing Contact Interactions at High Energy Lepton Colliders
\thanks{Work supported by NSERC (Canada) and the US DOE under contracts
DE-AC03-76SF00515 and DE-FG03-93ER40757.
}}
\medskip

{\large Kingman Cheung$^1$, Stephen Godfrey$^2$, and JoAnne Hewett$^3$} \\
\vskip .3cm
$^1$University of Texas, Austin TX 78712 \\
\vskip .3cm
$^2$Ottawa Carleton Institute for Physics, 
Carleton University, Ottawa, Canada\\
\vskip .3cm
$^3$Stanford Linear Accelerator Center, Stanford, CA 94309
\vskip .3cm

\end{center}

\begin{abstract} 

Fermion compositeness and other new physics
can be signalled by the presence of a strong 
four-fermion contact interaction.  Here we present 
a study of $\ell\ell qq$ and $\ell\ell\ell'\ell'$ contact interactions using 
the reactions: $\ell^+ \ell^- \to \ell'^+\ell'^-,b\bar b, c\bar c$ at future 
$e^+e^-$ linear colliders with $\sqrt s=0.5-5$ TeV and $\mu^+\mu^-$ 
colliders with $\sqrt s=0.5,4$ TeV.  We find that very large compositeness
scales can be probed at these machines and that the use of polarized beams
can unravel their underlying helicity structure. 

\end{abstract} 

\vskip0.75in
\noindent{To appear in the {\it Proceedings of the 1996 DPF/DPB Summer
Study on New Directions for High Energy Physics - Snowmass96}, Snowmass, CO,
25 June - 12 July, 1996.}

\renewcommand{\thefootnote}{\arabic{footnote}} \end{titlepage}

\section{Introduction}

There is a strong historical basis for the consideration of composite models 
which is presently mirrored in the proliferation of fundamental particles.
In attempts to explain the repetition of generations or the large number
of arbitrary parameters within the Standard Model (SM), several levels of 
substructure have been considered\cite{revs}, including composite 
fermions, Higgs bosons, and even weak bosons.
Here we focus on the possibility that leptons and quarks are bound states of
more fundamental constituents, often referred to as preons in the 
literature.
The preon binding force should be confining at a mass scale $\Lambda$ 
which also characterizes the radius of the bound states.  Experimentally, 
$\Lambda$ is constrained to be at least in the TeV range.  Theoretically,
numerous efforts have been made to construct realistic models for
composite fermions, but no consistent or compelling theory which accounts for
$m_{\ell,q}\ll\Lambda$ presently exists.  At energies above $\Lambda$ the
composite nature of fermions would be revealed by the break-up of the bound
states in hard scattering processes.  At lower energies, deviations from the
SM may be observed via form factors or residual effective interactions
induced by the binding force.  These composite remnants are usually
parameterized by the introduction of contact terms in the low-energy 
lagrangian.  More generally, four fermion contact interactions represent a 
useful parameterization of many types of new physics originating at a high energy
scale, such as the exchange of new gauge bosons, leptoquarks, or excited
particles, or the existence of anomalous couplings.

These contact interactions are described by non-renormalizable operators in
the effective low-energy lagrangian.  The lowest order four-fermion contact
terms are dimension-6 and hence have dimensionful coupling constants
proportional to $g^2_{eff}/\Lambda^2$.  The fermion currents are restricted
to be helicity conserving, flavor diagonal, and $SU(3) \otimes SU(2) \otimes 
U(1)$ invariant.  These terms can be written most generally as\cite{elp,lane}
\begin{equation}
{\cal L}  = \frac{g^2_{eff}\eta}{2\Lambda^2} \biggr( \bar q \gamma^\mu q + 
 {\cal F}_\ell \bar \ell \gamma^\mu \ell \biggr )_{L/R}
\; \biggr( \bar q \gamma_\mu q + 
 {\cal F}_\ell \bar \ell \gamma_\mu \ell \biggr )_{L/R} 
\end{equation}
where the generation and color indices have been suppressed, $\eta=\pm 1$, and
${\cal F}_\ell$ is inserted to allow for different quark and lepton couplings
but is anticipated to be ${\cal O}(1)$.   Since the binding force is expected
to be strong when $Q^2$ approaches $\Lambda^2$, it is conventional to define 
$g^2_{eff}=4\pi$.  The subscript $L/R$ indicates that
the currents in each parenthesis can be either left- or right-handed and
various possible choices of the chiralities lead to different predictions
for the angular distributions of the reactions where the contact terms
contribute.

Interference between the contact terms and the usual gauge interactions can
lead to observable deviations from SM predictions at energies lower than 
$\Lambda$.   They can affect {\it e.g.}, jet production at hadron colliders, 
the Drell-Yan process, or lepton scattering.  
The size of this interference term relative to the SM amplitude is 
$~ Q^2/\alpha_i \Lambda^2$, where $\alpha_i$ represents the strength of 
the relevant gauge coupling.  One may hence neglect modifications of the
gauge couplings due to form factors.  It is clear that the effects of the
contact interactions will be most important in the phase space region with 
large $Q^2$.  At hadron colliders these terms manifest themselves
in the high $E_T$ region in jet and lepton-pair production and deviations from
the SM can unfortunately often become entangled in the uncertainties associated
with the parton densities\cite{cdf-1}.  CDF has recently constrained\cite{cdf-2}
$qqqq$ contact interactions from the measurement of the dijet angular 
distribution; it is found to be in good agreement with QCD, thereby excluding 
at 95 \% C.L. a contact interaction among the 
up and down type quarks with scale $\Lambda^+ \le 1.6$ TeV or $\Lambda^-
\le 1.4$ TeV.  Composite scales of 2--5 TeV can be reached in future runs 
at the Tevatron\cite{cdf-2} and a search limit for $\Lambda$ in the 15--20 TeV 
range is expected at the LHC \cite{lhc}.  CDF also found the 
restrictions\cite{barbaro} $\Lambda^-_{LL} \ge 3.4$ TeV and 
$\Lambda_{LL}^+\ge 2.4$ TeV at 95\% C.L. on $qq\ell\ell$ contact interactions 
from $110\;{\rm pb}^{-1}$ of data on Drell-Yan production.  Run II of the 
Tevatron is expected to improve these limits to $~ 10$ TeV.  HERA also
constrains $qq\ell\ell$ contact terms, with the exclusion\cite{hera} from H1 
of $\Lambda\ge 1-2.5$ TeV at 95\% C.L., where the range takes into account 
various helicity combinations.  A review of the bounds on four lepton contact
interactions from PEP, PETRA, TRISTAN, and ALEPH is given in Buskulic
{\it et al.}\cite{aleph}, but are superseded by recent results from OPAL
at $\sqrt s = 161 $ GeV which bounds $\Lambda$ in the range $1.4-6.6$ TeV 
(again, for the various helicity states) from $e^+e^-, \mu^+\mu^-, \tau^+\tau^-$
and combined $\ell^+\ell^-$ pair production \cite{opal}.  
This search also constrains
$\Lambda_{eeqq}\ge 2.1-3.5$ TeV and $\Lambda_{eebb}\ge 1.6-3.7$ TeV at 95\%
C.L. from identified b-quark final states.  There is an earlier
analysis\cite{kaoru} of $eecc$
contact terms from the forward-backward asymmetry of $D$ and $D^*$ mesons
which yields a bound of $\Lambda_{eecc}>1 - 1.6$ TeV. 

In this contribution, we study the compositeness search reach on $\ell\ell qq$
and $\ell\ell\ell'\ell'$ contact terms using the processes 
$\ell^+\ell^- \to b\bar b,c\bar c, {\ell'}^+ {\ell'}^- $ where $\ell=e$ or
$\mu$ at future lepton colliders.   We shall consider $e^+e^-$ colliders
with center of mass energy 0.5, 1, 1.5, 5 TeV and luminosity 50, 200, 200, 
1000 fb$^{-1}$, respectively, as well as muon colliders with $\sqrt s=0.5, 4$ 
TeV and luminosity 0.7, 50, and 1000 fb$^{-1}$.  We build on earlier
studies\cite{dieter} of compositeness searches at lepton colliders.

\section{Collection of Formulae}

The reactions $\ell^+ \ell^- \to f\bar{f}$ where $f=\mu,\tau,b,c$ and 
$\ell=e,\mu$ ($\ell \not = f)$ proceed via $s$-channel exchanges of $\gamma$ 
and $Z$ bosons, as well as the $\ell\ell f \bar f$ contact interaction.
Thus, not only the squared term of the contact interaction but also the
interference terms between the $\gamma$, $Z$ exchanges and the contact 
interaction will contribute to the differential cross section and yield
deviations from the SM.
We explicitly rewrite the contact terms for $\ell\ell f\bar f$ 
\begin{eqnarray}
{\cal L} &=&{4\pi\over 2\Lambda^2} [ \eta_{LL} (\bar{e}_L \gamma_\mu e_L)
(\bar{f}_L \gamma^\mu f_L) \nonumber \\
& + &   \eta_{LR} (\bar{e}_L \gamma_\mu e_L) (\bar{f}_R \gamma^\mu f_R)
 +  \eta_{RL} (\bar{e}_R \gamma_\mu e_R) (\bar{f}_L \gamma^\mu f_L)
\nonumber \\
& + & \eta_{RR} (\bar{e}_R \gamma_\mu e_R) (\bar{f}_R \gamma^\mu f_R) ] \,.
\end{eqnarray}
The polarized differential cross sections for $e^-_{L/R} e^+ \to f\bar f$
are 
\begin{equation}
{{d\sigma_L} \over {d\cos\theta}} = {{\pi \alpha^2 C_f}\over{4s}}
\left\{ |C_{LL}|^2(1+\cos\theta)^2 +|C_{LR}|^2 (1-\cos\theta)^2 \right\} 
\end{equation}
\begin{equation}
{{d\sigma_R} \over {d\cos\theta}} = {{\pi \alpha^2 C_f}\over{4s}}
\left\{ |C_{RR}|^2(1+\cos\theta)^2 +|C_{RL}|^2 (1-\cos\theta)^2 \right\}
\end{equation}
where $\theta$ is the scattering angle in the center of mass frame and $C_f$
represents the color factor being the usual 3(1) for quarks(leptons).  The
helicity amplitudes are
\begin{eqnarray}
C_{LL} &=& -Q_f +{{C_L^e C_L^f}\over {c_w^2 s_w^2}} 
{s\over{s-M_Z^2+i\Gamma_Z M_Z}} 
+{{s\eta_{LL}}\over{2\alpha\Lambda^2}} \nonumber\,,\\
C_{LR} &=& -Q_f +{{C_L^e C_R^f}\over {c_w^2 s_w^2}} 
{s\over{s-M_Z^2+i\Gamma_Z M_Z}} 
+{{s\eta_{LR}}\over{2\alpha\Lambda^2}} \,,
\end{eqnarray}
with $C_L^f=T_{3f} - Q_f s_w^2$, $C_R^f= - Q_f s_w^2$, $s_w$ and $c_w$ 
are the sine and cosine of the weak mixing angle, and $Q_f$ and $T_{3f}$ 
represent the fermion's charge and third component of the weak isospin, 
respectively.
The expressions for $C_{RR}$ and $C_{RL}$ can be obtained by 
interchanging $L \leftrightarrow R$.  The use of polarized beams, combined
with the angular distributions, can thus clearly determine the helicity
of the contact term.

The unpolarized differential cross section is given simply by
\begin{equation}
{{d\sigma} \over {d\cos\theta}} ={1\over 2} 
\left[{{d\sigma}_L \over {d\cos\theta}}+{{d\sigma}_R \over {d\cos\theta}}
\right]\,.
\end{equation}
The polarized and unpolarized total cross sections are obtained by
integrating over $\cos\theta$, resulting in the spin-averaged unpolarized 
cross section:
\begin{equation}
\sigma = {{\pi \alpha^2 C_f}\over {3s}} \biggr[ |C_{LL}|^2 + |C_{LR}|^2
+ |C_{RL}|^2 + |C_{RR}|^2 \biggr]\,.
\end{equation}
The forward-backward and left-right asymmetries are easily obtained and
can be written as 
\begin{eqnarray}
A_{FB} &=& 
\frac{3}{4}\, \frac{|C_{LL}|^2 + |C_{RR}|^2 - |C_{LR}|^2 -|C_{RL}|^2}
                       {|C_{LL}|^2 + |C_{RR}|^2 + |C_{LR}|^2 +|C_{RL}|^2}\,,
\\
A_{LR} &=&  \frac{|C_{LL}|^2 + |C_{LR}|^2 - |C_{RR}|^2 -|C_{RL}|^2}
                 {|C_{LL}|^2 + |C_{LR}|^2 + |C_{RR}|^2 +|C_{RL}|^2}\,.
\end{eqnarray}
Figure~1 displays the $\cos\theta$ distribution for $e^+e^-
\to b\bar b$ at $\sqrt{s}=0.5$ TeV for the SM and with a contact term 
present.  The effects of a contact term are qualitatively similar for 
other final states.
In all curves we set $|\eta|=1$. In fig. 1a we take 
$\Lambda=10$~TeV which shows that a finite value of $\Lambda$ alters the 
angular distribution particularly in the forward direction.  Fig. 
1b  displays the angular distributions for right-handed polarized 
electrons.  Although the effects of contact interactions are 
more dramatic here, because the right-handed cross section is 
smaller, the relative contribution of right-handed contact terms will be 
smaller in unpolarized cross sections.  Thus,  not only will 
polarization be important for disentangling the helicity structure of 
a contact interaction should deviations be seen, but polarization will 
also enhance the sensitivity to contact interactions.  In fig. 1c 
distributions are shown for $\eta_{LL}=\pm 1$ and $\eta_{RR}=\pm 1$ 
demonstrating that opposite signs for the $\eta$'s results in opposite 
interference.  Finally, in fig. 1d the angular distribution is shown 
for $\eta_{LL}$ but with $\Lambda=$ 5~TeV, 10~TeV, 20~TeV, and 30~TeV 
to give a feeling for the sensitivity to the scale of new physics.

\begin{figure}[t]
\leavevmode
\centerline{
\epsfig{file=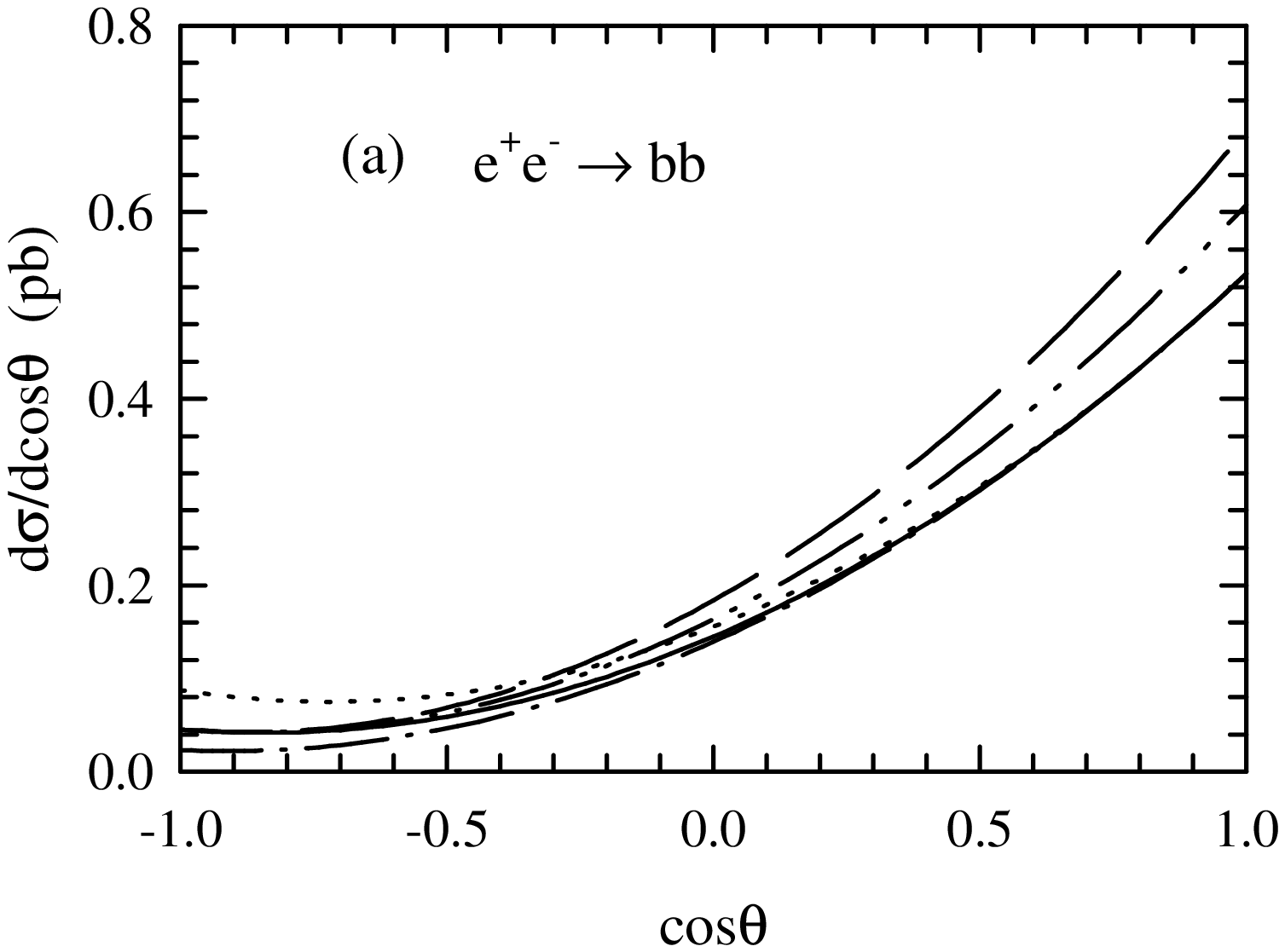,height=8.0cm,width=8.0cm,clip=}
\hspace*{-5mm}
\epsfig{file=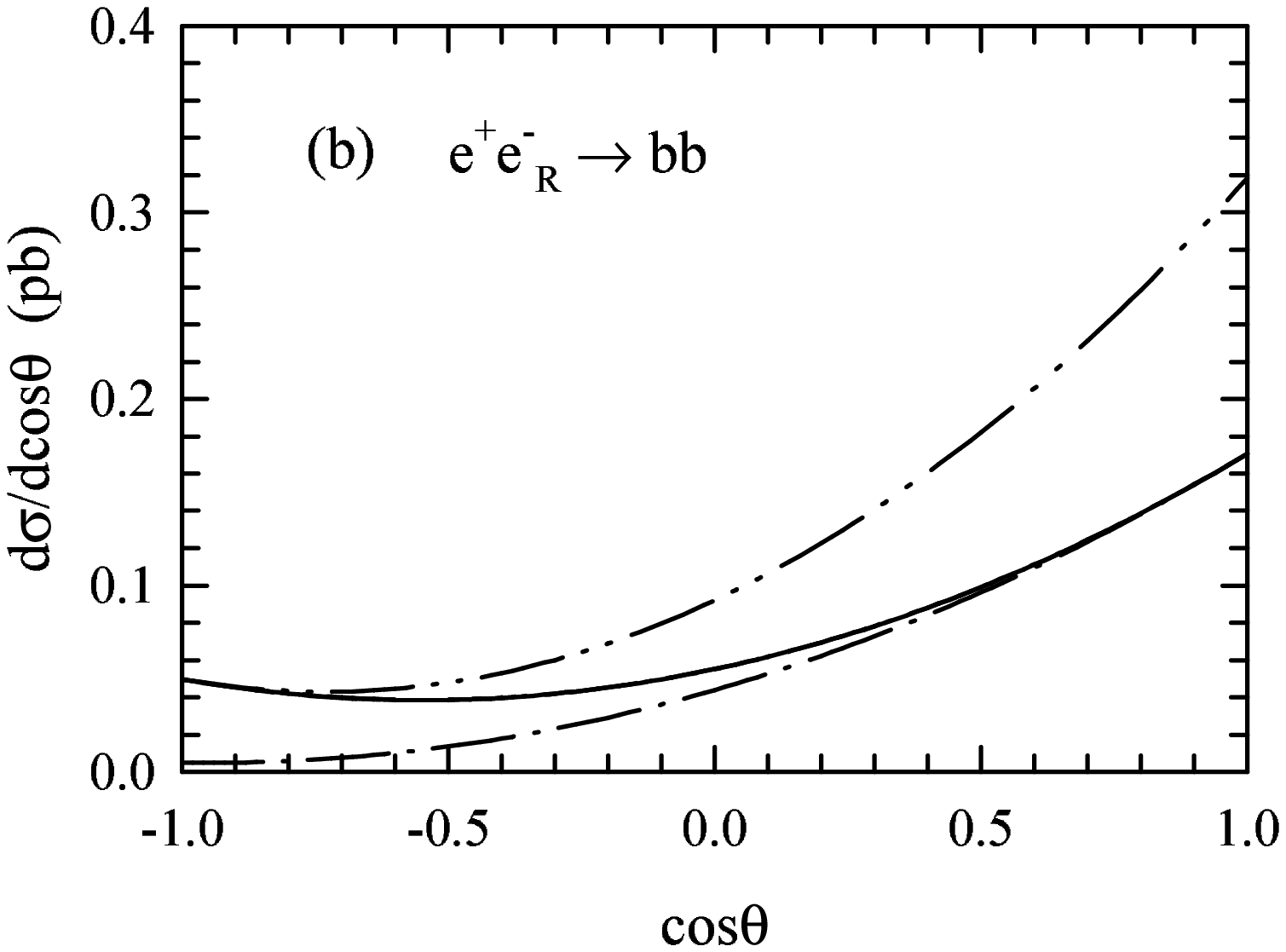,height=8.0cm,width=8.0cm,clip=}}
\vspace*{-1.cm}
\centerline{
\epsfig{file=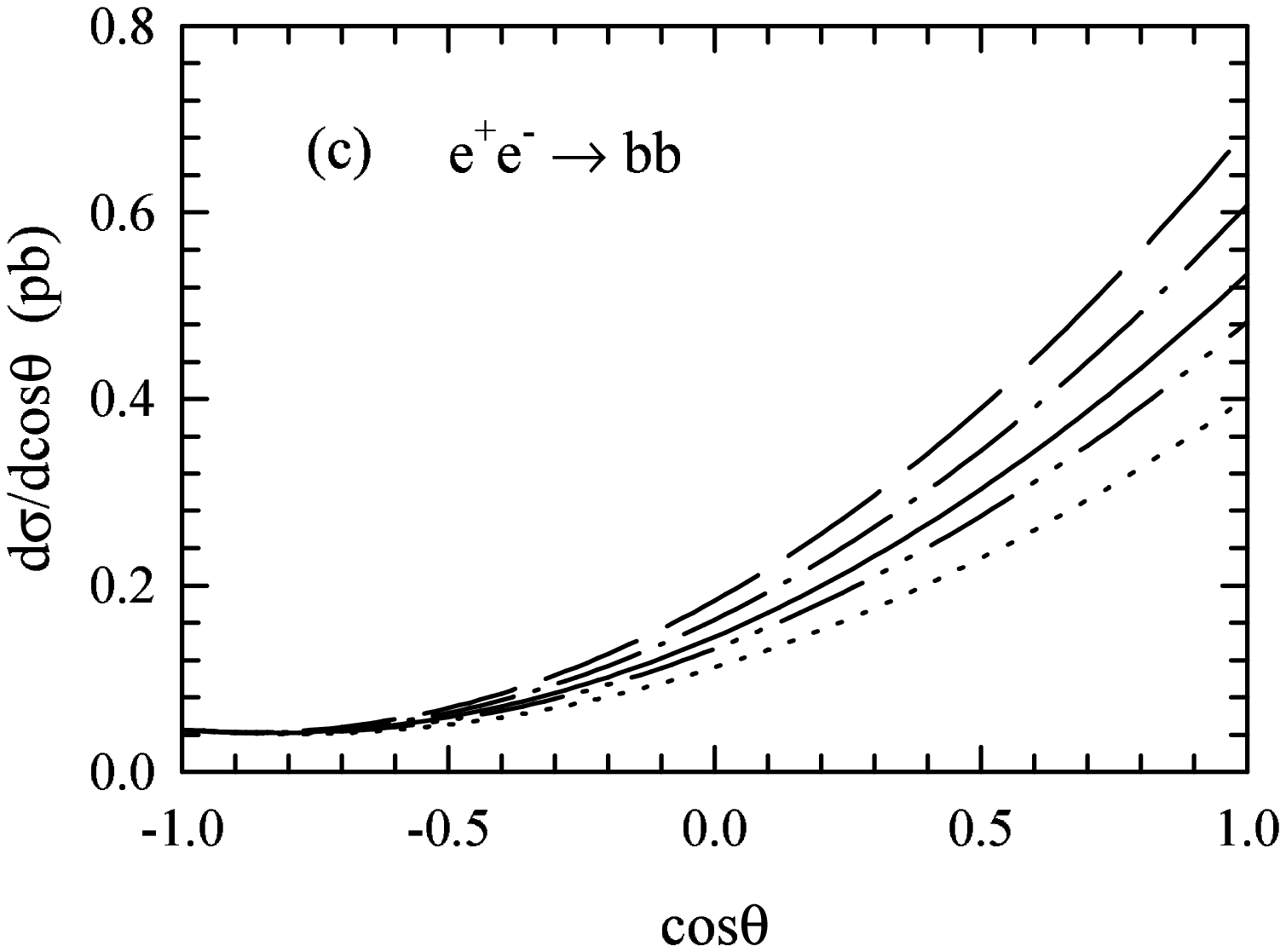,height=8.0cm,width=8.0cm,clip=}
\hspace*{-5mm}
\epsfig{file=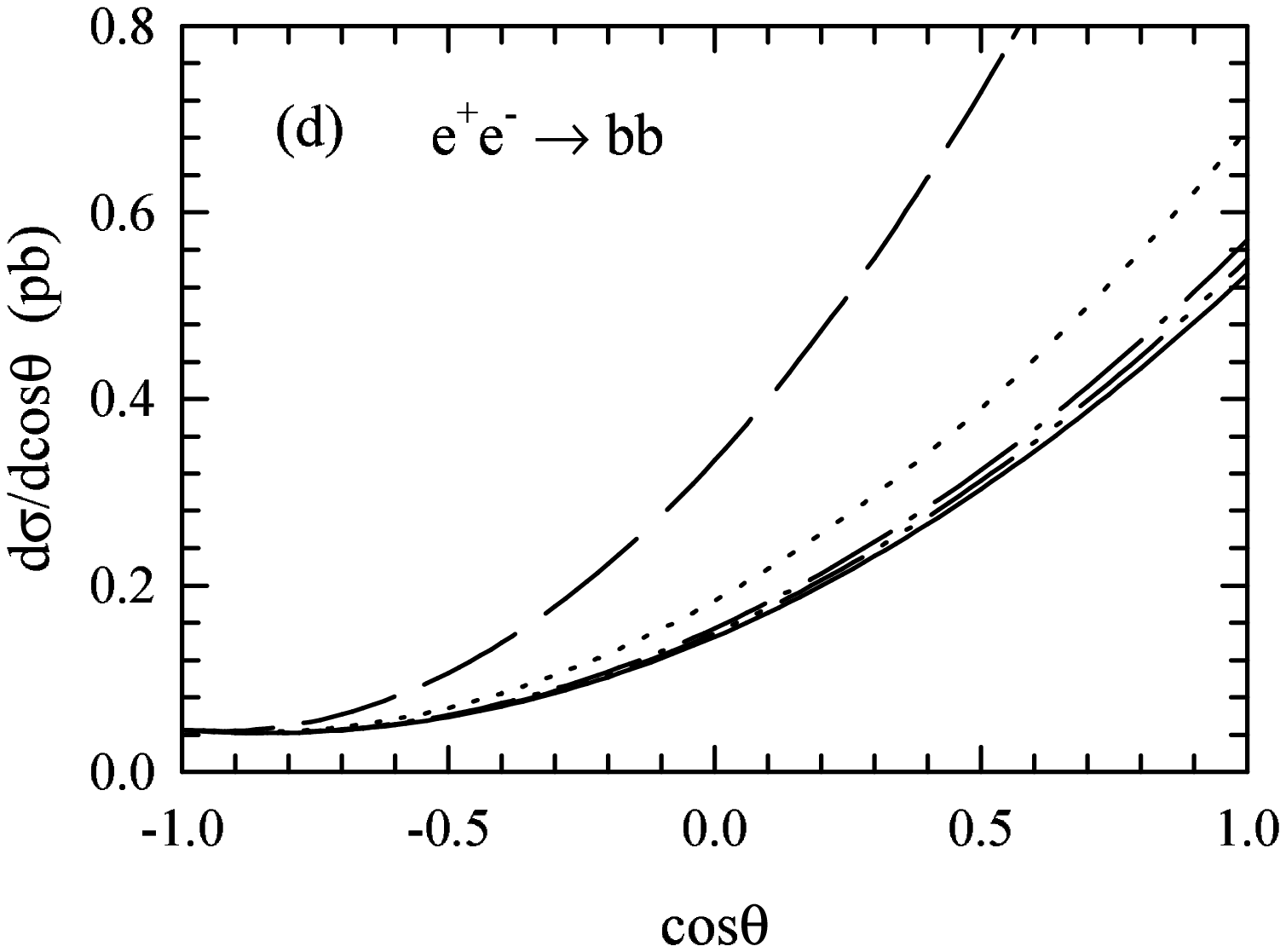,height=8.0cm,width=8.0cm,clip=}}
\vspace*{0.5cm}
\caption{The $\cos\theta$ distribution for $e^+e^- \to b\bar{b}$ at 
$E_{CM} =0.5$ TeV
with $\Lambda= 10 $~TeV everywhere except for (d).
In all cases the solid line is for the SM ($\Lambda= \infty $).
(a) Unpolarized $e^+e^-$ with
$\eta_{LL}=+1$ (dashed), $\eta_{LR}=+1$ (dotted),
$\eta_{RL}=+1$ (dot-dashed), and $\eta_{RR}=+1$ (dot-dot-dashed).
(b) Polarized $e^+e^-_R$ with
$\eta_{RL}=+1$ (dot-dashed ) and $\eta_{RR}=+1$ (dot-dot-dashed).
(c) Unpolarized $e^+e^-$ with
$\eta_{LL}=+1,-1$ (dashed, dotted), 
$\eta_{RR}=+1,-1$ (dot-dashed, dot-dot-dashed).
(d) Unpolarized $e^+e^-$ with $\eta_{LL}=+1$ with
$\Lambda= 5, 10, 20, 30 $~TeV (dashed, dotted, dot-dashed,
dot-dot-dashed).}
\end{figure}

We note that the effects of the contact term on $e^+e^- \to q\bar q$
are relatively small when all quark flavors are summed,
compared to the individual deviations in {\it e.g.}, $b\bar b$
or $c\bar c$, because cancelations occur in the interference term 
between the up-type and 
down-type quarks.  We thus concentrate on the heavy quark final states,
taking a 60\% identification efficiency for detecting b-quarks and 35\% 
identification efficiency for detecting c-quarks at the NLC\cite{davej}.
The detection efficiency of heavy flavor final states at a muon collider
has yet to be determined, but is expected to be worse than what can be
achieved at the NLC due to the inability to put a vertex detector close to
the interaction point and due to the heavier backgrounds.  For now we 
assume canonical LEP values, $\epsilon_b=25\%, \epsilon_c=5\%$ for the muon
collider but warn the reader 
that these numbers are quite arbitrary and are only used for 
illustrative purposes.
We assume 100\% identification efficiency for leptons.
Although we do not take into account the purity of the 
tagged heavy flavor samples in our results, we note that the purities 
that can be achieved at a linear collider are higher than can be 
achieved at LEP.

\section{Results}

To gauge the  sensitivity to the compositeness scale
we assume that the SM is correct and perform a $\chi^2$ analysis of the
$\cos\theta$ angular distribution.  
To perform this we choose the detector acceptance to be 
$|\cos\theta|<0.985$  (corresponding to $\theta=10^o$)
for the $e^+e^-$ collider 
and $|\cos\theta|<0.94$ (corresponding to $\theta=20^o$)
for the muon collider\cite{nlcmuon}.
We note that angular acceptance of a typical muon collider detector is expected
to be reduced due to additional shielding required to minimize the radiation 
backgrounds from the muon beams.  We then divide 
the angular distribution into 10 equal bins.
The $\chi^2$ distribution is evaluated by the usual expression:
\begin{equation}
\chi^2 = L\times\epsilon\times\sum_{i=1}^{10} 
\left[ {
{\int_{\hbox{bin i}} {{d\sigma^\Lambda }\over{d\cos\theta}} d\cos\theta
-\int_{\hbox{bin i}} {{d\sigma^{SM} }\over{d\cos\theta}} d\cos\theta }
\over
{\sqrt{ \int_{\hbox{bin i}} { {d\sigma^{SM}} \over{d\cos\theta} } 
d\cos\theta} } 
}\right]^2
\end{equation}
where $L$ is the luminosity and $\epsilon$ is the efficiency for detecting 
the final state under consideration which is discussed above.  For 
polarized beams we assume $1/2$ of the total integrated luminosity 
listed in the tables for each polarization.
We assume that only one of the $\eta$'s is nonzero at a time.

The 95\% C.L. bounds on $\Lambda$ are tabulated in Tables I, II, and 
III\footnote{The limits in this paper supersede the results presented in
the New Interaction Subgroup Report by K. Cheung and R. Harris\cite{rob}}.
Generally, high luminosity $e^+ e^-$ and $\mu^+\mu^-$ colliders 
are quite sensitive to contact interactions with discovery limits 
ranging from 5 to 50 times the center of mass energy.  For 
unpolarized beams with leptons in the final state,
the slightly higher sensitivity to contact interactions at
$e^+ e^-$ colliders than at $\mu^+\mu^-$ colliders with the same 
$\sqrt{s}$ can be attributed 
to the larger expected acceptance for $e^+e^-$ detectors.  For 
$b\bar{b}$ final states the sensitivities for $e^+ e^-$ colliders are 
roughly 20\% higher than for $\mu^+ \mu^-$ colliders while for 
$c\bar{c}$ final states the difference can be up to a factor of two.  
These differences are due to the different tagging efficiencies 
assumed for $e^+e^-$ and $\mu^+\mu^-$ colliders.
Polarization in the $e^+e^-$ colliders can offer even higher limits 
depending on the final state being considered.  More importantly,  if 
deviations are observed polarization would be crucial for determining 
the chirality of the new interaction.  Finally, we note that for 
$b\bar{b}$ and $c\bar{c}$ final states sometimes very specific, relatively 
low values of $\Lambda$, give rise to angular
distributions indistinguisiable 
from the SM.  However, we expect that these values will be ruled out 
by other measurements before high energy lepton colliders become 
operational so we only include the higher values in the tables.

\section{Summary}

In this report we presented the results of a preliminary study on the 
sensitivity to contact interactions at future high energy 
$e^+ e^-$ and $\mu^+\mu^-$ colliders.  Depending on the specific 
collider and final state, contact interactions can be detected up to 
5-50 times the center of mass energy of the collider with the lowest 
number coming from the low luminosity 500~GeV $\mu^+\mu^-$ collider 
and the highest numbers from high luminosity $e^+e^-$ 
colliders with polarization. These results should be taken as 
preliminary.  First and foremost the sensitivities were based only on 
statistical errors and systematic errors were not included.  In 
addition, a more thorough analysis should include potentially 
important effects like initial state radiation and should consider 
heavy quark final state purities.  These considerations are under 
study and will be presented elsewhere \cite{paper}.

\section*{Acknowledgements}

S.G. thanks Dean Karlen for helpful conversations.

\begin{table}
\begin{center}
{\small
\caption{95\% C.L. compositeness search reach in TeV for $e^+e^-$ colliders.}
\begin{tabular}{lllll}
\hline
\hline
 process &
$\Lambda_{LL}$ &  $\Lambda_{LR}$ & $\Lambda_{RL}$ & $\Lambda_{RR}$  \\ \hline 
\multicolumn{5}{c}{$\sqrt{s}=0.5$ TeV,  $L$=50 fb$^{-1}$} \\ \hline
 $e^-_L e^+ \to \mu^+ \mu^-$ (P=1.0) & 33 & 30 & --- & --- \\
 $e^-_L e^+ \to \mu^+ \mu^-$ (P=0.9) & 32 & 29 & 10 & 10 \\
 $e^-_R e^+ \to \mu^+ \mu^-$ (P=1.0) &  --- &  --- &   30 &   33 \\
 $e^-_R e^+ \to \mu^+ \mu^-$ (P=0.9) & 11 & 10 &   29 &   31 \\
 $e^- e^+ \to \mu^+ \mu^-$ & 28 & 26 &   26 & 27 \\
 $e^-_L e^+ \to b \bar{b}$ (P=1.0) & 39 & 32 &    --- &    --- \\
 $e^-_L e^+ \to b \bar{b}$ (P=0.9) & 38 & 30 &  5.3 &  9.1 \\
 $e^-_R e^+ \to b \bar{b}$ (P=1.0) &  --- &  --- &    35 &   38 \\
 $e^-_R e^+ \to b \bar{b}$ (P=0.9) & 17 & 12 & 33 & 33 \\
 $e^- e^+ \to b \bar{b}$ & 37 & 28 & 28 & 25 \\
 $e^-_L e^+ \to c \bar{c}$ (P=1.0) & 34 & 29 &    --- &    --- \\
 $e^-_L e^+ \to c \bar{c}$ (P=0.9) & 33 &  28 &    5.0 &    8.5 \\
 $e^-_R e^+ \to c \bar{c}$ (P=1.0) &  --- &  --- &    27 &   34 \\
 $e^-_R e^+ \to c \bar{c}$ (P=0.9) &  12 &  11 &    24 &   32 \\
 $e^- e^+ \to c \bar{c}$ & 31 &  28 &    18 &   26 \\ 
\hline
\multicolumn{5}{c}{$\sqrt{s}=1$ TeV,  $L$=200 fb$^{-1}$} \\ \hline
 $e^-_L e^+ \to \mu^+ \mu^-$ (P=1.0) & 66 & 60 & --- & --- \\
 $e^-_L e^+ \to \mu^+ \mu^-$ (P=0.9) &   63 & 57 &   20 &   21 \\
 $e^-_R e^+ \to \mu^+ \mu^-$ (P=1.0) &    --- &  --- &   61 &   66 \\
 $e^-_R e^+ \to \mu^+ \mu^-$ (P=0.9) &   22 & 20 &   58 &   62 \\
 $e^- e^+ \to \mu^+ \mu^-$ &   57 & 51 &   51 &   55 \\
 $e^-_L e^+ \to b \bar{b}$ (P=1.0) &   78 & 64 &    --- &    --- \\
 $e^-_L e^+ \to b \bar{b}$ (P=0.9) &   75 & 61 &   11 &   18 \\
 $e^-_R e^+ \to b \bar{b}$ (P=1.0) &   --- &  --- &   70 &   76 \\
 $e^-_R e^+ \to b \bar{b}$ (P=0.9) &  34 & 24 &   65 &   67 \\
 $e^- e^+ \to b \bar{b}$ &  74 & 56 &   55 &   50 \\
 $e^-_L e^+ \to c \bar{c}$ (P=1.0) &  68 & 59 &    --- &    --- \\
 $e^-_L e^+ \to c \bar{c}$ (P=0.9) &  65 & 57 &    9.9 &   17 \\
 $e^-_R e^+ \to c \bar{c}$ (P=1.0) &   --- &  --- &  55 &   68 \\
 $e^-_R e^+ \to c \bar{c}$ (P=0.9) &  24 &  22 &   49 & 62 \\
 $e^- e^+ \to c \bar{c}$ &  61 &  55 &   37 & 51 \\
\hline \hline
\end{tabular}}
\end{center}
\end{table}

\begin{table}
\begin{center}
{\small
\caption{95\% C.L. compositeness search reach in TeV for $e^+e^-$ colliders.}
\begin{tabular}{lllll}
\hline
\hline
process &
$\Lambda_{LL}$ &  $\Lambda_{LR}$ & $\Lambda_{RL}$ & $\Lambda_{RR}$  \\ \hline 
\multicolumn{5}{c}{$\sqrt{s}=1.5$ TeV,  $L$=200 fb$^{-1}$} \\ \hline
 $e^-_L e^+ \to \mu^+ \mu^-$ (P=1.0) & 81 &   74 & --- & --- \\
 $e^-_L e^+ \to \mu^+ \mu^-$ (P=0.9) &   77 & 70 &   25 &   26 \\
 $e^-_R e^+ \to \mu^+ \mu^-$ (P=1.0) &    --- &  --- &   74 &   81 \\
 $e^-_R e^+ \to \mu^+ \mu^-$ (P=0.9) &   27 & 25 &   70 &   76 \\
 $e^- e^+ \to \mu^+ \mu^-$ &   70 & 63 &   63 &   67 \\
 $e^-_L e^+ \to b \bar{b}$ (P=1.0) &   95 & 80 &    --- &    --- \\
 $e^-_L e^+ \to b \bar{b}$ (P=0.9) &   92 & 77 &   15 &   23 \\
 $e^-_R e^+ \to b \bar{b}$ (P=1.0) &   --- &  --- &   84 &   94 \\
 $e^-_R e^+ \to b \bar{b}$ (P=0.9) &  41 & 31 &   78 &   82 \\
 $e^- e^+ \to b \bar{b}$ &  90 & 70 &   66 &   61 \\
 $e^-_L e^+ \to c \bar{c}$ (P=1.0) &  83 & 72 &    --- &    --- \\
 $e^-_L e^+ \to c \bar{c}$ (P=0.9) &  80 & 69 &   14 &    20 \\
 $e^-_R e^+ \to c \bar{c}$ (P=1.0) &   --- &  --- &   67 &    83 \\
 $e^-_R e^+ \to c \bar{c}$ (P=0.9) &   29 & 26 &   60 &    76 \\
 $e^- e^+ \to c \bar{c}$ &  75 & 67 &  44 &    63 \\ 
\hline
\multicolumn{5}{c}{$\sqrt{s}=5$ TeV,  $L$=1000 fb$^{-1}$} \\ \hline
 $e^-_L e^+ \to \mu^+ \mu^-$ (P=1.0) & 220 & 200 & --- & --- \\
 $e^-_L e^+ \to \mu^+ \mu^-$ (P=0.9) &  210 & 190 &  70 &   71 \\
 $e^-_R e^+ \to \mu^+ \mu^-$ (P=1.0) &    --- &   --- & 200 &  220 \\
 $e^-_R e^+ \to \mu^+ \mu^-$ (P=0.9) &   75 &  70 & 190 &  210 \\
 $e^- e^+ \to \mu^+ \mu^-$ &  190 & 170 & 170 &  180 \\
 $e^-_L e^+ \to b \bar{b}$ (P=1.0) &  260 & 220 &   --- &    --- \\
 $e^-_L e^+ \to b \bar{b}$ (P=0.9) &  250 & 210 &  49 &   66 \\
 $e^-_R e^+ \to b \bar{b}$ (P=1.0) &   --- &   --- &  230 &  250 \\
 $e^-_R e^+ \to b \bar{b}$ (P=0.9) & 110 &  89 &  210 &  220 \\
 $e^- e^+ \to b \bar{b}$ & 250 & 200 &  180 &  170 \\
 $e^-_L e^+ \to c \bar{c}$ (P=1.0) & 220 & 190 &   --- &    --- \\
 $e^-_L e^+ \to c \bar{c}$ (P=0.9) & 210 & 190 &  43 & 46 \\
 $e^-_R e^+ \to c \bar{c}$ (P=1.0) &   --- &   --- & 180 & 220 \\
 $e^-_R e^+ \to c \bar{c}$ (P=0.9) & 78 &  38 & 160 & 210 \\
 $e^- e^+ \to c \bar{c}$ & 200 & 180 & 110 & 170 \\
\hline \hline
\end{tabular}}
\end{center}
\end{table}

\begin{table}
\begin{center}
\caption{95\% C.L. compositeness search reach in TeV for 
$\mu^+\mu^-$ colliders.}
\begin{tabular}{lllll}
\hline
\hline
 process &
$\Lambda_{LL}$ &  $\Lambda_{LR}$ & $\Lambda_{RL}$ & $\Lambda_{RR}$  \\ \hline 
\multicolumn{5}{c}{$\sqrt{s}=0.5$ TeV,  $L$=0.7 fb$^{-1}$} \\ \hline
$\mu^+ \mu^- \to \tau^+ \tau^-$ &    9.8 &  9.1 & 9.1 & 9.4 \\
$\mu^+ \mu^- \to b \bar{b}$ & 10 & 8.8 & 4.9 & 7.6 \\
$\mu^+ \mu^- \to c \bar{c}$ & 5.6 & 3.6 & 4.2 & 2.7 \\
\hline
\multicolumn{5}{c}{$\sqrt{s}=0.5$ TeV,  $L$=50 fb$^{-1}$} \\ \hline
 $\mu^+ \mu^- \to \tau^+ \tau^-$ &   28 &   25 &   25 &   27 \\
 $\mu^+ \mu^- \to b \bar{b}$ &  29 &   22 & 21 &   20 \\
 $\mu^+ \mu^- \to c \bar{c}$ &  19 & 16 &    5.7 & 15 \\
\hline
\multicolumn{5}{c}{$\sqrt{s}=4$ TeV,  $L$=1000 fb$^{-1}$} \\ \hline
 $\mu^+ \mu^- \to \tau^+ \tau^-$ &  170 &  150 &  150 &  160 \\
 $\mu^+ \mu^- \to b \bar{b}$ & 180 & 140 & 120 & 120 \\
 $\mu^+ \mu^- \to c \bar{c}$ & 110 &  92 & 42 & 90 \\
\hline \hline
\end{tabular}
\end{center}
\end{table}


\end{document}